# Interactional Expertise and Embodiment

## Harry Collins

### For

## Abstract

In Part 1 of this paper, I introduce the idea of interactional expertise while in Part 2, I focus on its implications for philosophical theories of the importance of the body in forming our conceptual world. I argue that the way philosophers have dealt with the body turns attention away from the most important questions and that we cannot answer these questions without making the notion of socialisation, and therefore interactional expertise, a central concept in our thinking. This makes language at least as important, and often more important than bodily practice in our understanding of the world. The notion of a disembodied socialised agent leads in the direction of interesting questions while the notion of an embodied but unsocialised human actor is unimaginable.

## Part I: Interactional Expertise

The philosophical *idea* of interactional expertise first arose before the *term* was invented. This was in the mid-1990s, in the context of the discussion of the limitations of artificial intelligence (AI); the question was can machines without human-like bodies be intelligent?[1] The first published appearance of *the term* 'interactional expertise' (IE) was in the 'Third Wave' paper by Collins and Evans, published in 2002 but this paper dealt with the concept 'by-the-way' while attempting to shift social scientists' attention to expertise in general. The first full discussion of the term is found in a 2004(a) paper entitled 'Interactional Expertise as a Third Kind of Knowledge' which draws together the AI stream of thinking, fieldwork observations and an analysis of language. In all, four channels feed into the *idea* of interactional expertise, as shown in Table 1. The backbone of the concept is the philosophical stream. The 2004a article unites the philosophy stream with the fieldwork 'tributary' and an imitation game 'tributary'. There is also a 'sociology and policy' channel which, on pain of some convoluted hydraulics, is mostly downstream from the others in a sideways kind of way.[2]

---

[1] Collins 1996a, the original source, was a review of Hubert Dreyfus's *What Computers Still Can't Do*.

[2] This Figure, and much of Part 1, is adapted from Collins and Evans 2015 which deals with each of the entries in much greater detail than here.

| IMITATION GAME | PHILOSOPHY | FIELDWORK | SOCIOLOGY POLICY |
|---|---|---|---|
| Winch 1958 Wittgenstein 1953 The sociological interpretation of Wittgenstein *1974 1975 ...* | | | |
| *1990* *Artificial Experts* Pre-IE experiments | *1996* **SEPARATION PRINCIPLE** Separate language and practice Embodiment | *2002* *The term 'interactional expertise'* | Epstein Sheep farmers |
| **2004** Colour blind, perfect pitch experiments begin | | *2004* *Gravity's Shadow* | Imagined problem of GW and power lines! Managers |
| *2004* **THE SYNTHESIS** | | | Peer reviewers Committees |
| **2005** Collins as GW physicist | *2007* Strong Interactional Hypothesis | | Sports coaches etc |
| **2007** The blind | | | Crossing linguistic divides |
| **2007** Classroom Experiments on Gays etc | | | Ambassadorial model |
| **2011** Large Scale Experiments begin | | | *2011* Special IE Division of Labour |
| | | *2013* *Gravity's Ghost and Big Dog* | *2015* Ubiquitous IE Social 'glue' |
| | | | *2016/17* The Owls |

*Table 1: Evolution of Idea of interactional expertise (dates of publications in italics)*

*The sociological interpretation of Wittgenstein*

The basic ideas that led to the concept of interactional expertise can be dated to the interpretation of Wittgenstein that gave rise to the sociology of scientific knowledge (as represented in the full-width black box at the head of Table 1). In our case, the start was a reading of Winch's (1958/1988) *The Idea of a Social Science* – a book which presents itself as a philosophical critique of sociology. The reading of Winch with which we started, 'stands Winch on his head' or, at least, 'on his side'. What is treated as crucial is Winch's argument that social and conceptual life are two sides of the same coin but the reading draws out the sociological implications of the idea instead of presenting it as a philosophical critique. In so far as a critique can be drawn from Winch's argument, philosophy is as much sociology as sociology is philosophy. This reading gives rise to what we will call 'the sociological interpretation Wittgenstein' and the consequent use of the work of the later Wittgenstein is similar to that recommended by Bloor.[3] This interpretation of Wittgenstein has been accepted by only a minority of philosophers but has proved fruitful in the sociology of scientific knowledge and other domains. The Winch-inspired sociological interpretation of Wittgenstein has given the diverse streams of the idea of interactional expertise a unity though the streams were not brought together in print, until 2004.

*The separation principle*

The sociological interpretation of Wittgenstein treats 'forms-of-life' in a Durkheimian way such that a social group's patterns of language and practice give rise to their understanding of the world. You cannot understand peoples' actions without understanding their concepts and you cannot understand their concepts without understanding their actions. Around page 120 of his little book, Winch explains this in terms of the hygiene-related actions of surgeons in an operating theatre and their integral dependency on the idea of 'germ'. *Inter alia* he invents the notion of scientific paradigm revolution four years before Kuhn's '*Structure …*' which was published in 1962.[4] The new thought that gives rise to interactional expertise is that the contribution to a form-of-life of language on the one hand, and practice on the other, can be analytically and empirically separated – this is what we are now calling the 'separation principle'. Under this model, language and practice *together* give rise to forms-of-life at the collective level but individuals can acquire a complete 'understanding' of a form-of-life through immersion in the *language alone* and without taking part in the practices. The principle, then, separates collectivities of humans from individuals; language and practice are inseparable at the collective level while separable at the individual level.[5] The formulation below, upon which I cannot improve, is taken from the 1996(a) publication referred to above:

> Wittgenstein said that if a lion could speak we would not understand it. The reason we would not understand it is that the world of a talking lion - its `form of life' - would be different from ours. Bringing back Dreyfus's chair example, lions would not have chairs in their language in the way we do because lions' knees do not bend as ours do, nor do lions `write, go to conferences or give lectures'. ... But this does not mean that every entity that can recognise a chair has to be able to sit on one. That confuses the capabilities of an individual with the form of life of the social group in which that individual is embedded. Entities that can recognise chairs have only to share the form of life of those who can sit down. We would not understand what a talking lion said to us, not because it had a lion-like body, but because the large majority of its friends and acquaintances had lion-like bodies and lion-like interests. In principle, if one could find a lion cub that had the potential to have conversations, one could bring it up in human society to speak about chairs as we do in spite of its funny legs. It would learn to recognise chairs as it learned to speak our language. This is how the Madeleine case is to be understood; Madeleine has undergone linguistic socialization. [Madeleine is a woman, discussed by Oliver Sacks (1985/2011), who was almost totally disabled from birth yet, according to Sacks, was completely fluent in spoken language.] In sum, the shape of the bodies of the members of a social collectivity and the situations in which they find themselves give rise to their form of life. Collectivities whose members have different bodies and encounter different situations develop different forms of life. But given the capacity for linguistic socialisation, an individual can come to share a form of life without having a body or the experience of physical situations which correspond to that form of life.

---

[3] The position finds its most explicit formulation in David Bloor's (1983) argument that the later (e.g. 1953/1958) Wittgenstein is to be thought of as a sociologist as much as a philosopher. See also Bloor (1976/1991)

---

[4] It gives rise to a very natural sociological/philosophical interpretation of the Kuhnian notion of paradigm (Kuhn 1962)

[5] For my own claim about why bodies are important at the collective level such that we cannot envisage a society of brains in vats or one mediated solely by electronic media without face-to-face contact, see Collins, 1996b. This argument, however, does not mean that we could not individual brains in vats which were connected into society. To show that is not possible, if it is not, requires a more difficult kind of argument and one can see how difficult it is going to be by noting the success of contemporary AI device which are plugged into society via Google-type internet connections..

I found myself reviewing this book because of my work on AI which developed from the mid-1980s. In this work I end up challenging Dreyfus's well-known claim that computers cannot act 'intelligently' because they do not have bodies.[6] My counter-claim is that computers are not intelligent, not because they don't have bodies but because we do not know how to embed them in society.  The example of Madeleine was put forward by AI-pioneer, Douglas Lenat, one of Dreyfus's critics, in order to show that one can learn language without having a human-like body and Collins agreed with Lenat on this point while still holding that computers cannot be intelligent because they cannot be socialised.  *The term* 'interactional expertise' was not used in these early contributions but the idea that language and practice must be separated if we are to be clear about the different relationship of the collectivity and the individual to a form-of-life is the key and fully expressed in the quoted passage.  The idea depends on the additional idea, worked out more carefully in the 2004a publication, that language is not formal, 'propositional knowledge', as it is treated by those keen to stress the practice element of forms-of-life, but is itself a tacit knowledge-laden form of social interaction.

---

[6] The first foray is found at the end of Chapter 1 of my 1985/1992 and it was followed by a prize-winning paper at the annual meeting of the British Computer Society Specialist Group on Expert Systems which in turn encouraged me to write my 1990. I knew nothing of Bert Dreyfus's work when I started this. If I had I

Philosophically, interactional expertise as it first appeared makes a new claim about how forms-of-life work and how individuals fit into them: mostly people in a society come to understand each other through socialisation that involves sharing both practices and language but sometimes this understanding is via acquisition of the spoken language alone.  It is argued that only this way can we understand, for example, how those who are sufficiently congenitally challenged to be unable to engage in common practices with others can come to participate in the common society by immersion in the spoken discourse alone.

---

probably would not have started because his work is so terrifyingly good but, as it is, I took a different and much more sociological line.

### *A bold conjecture: The strong interactional hypothesis*

The last entry in the philosophy stream in Table 1 is the 'strong interactional hypothesis' (SIH). It takes forward the idea of interactional expertise in the form of, to use Popper's (1959/2002) term, a 'bold conjecture':

> In principle, the level of fluency in the language of a domain that can be attained by someone who is an interactional expert only is indistinguishable from that which can be attained by a full-blown contributory expert in any test involving language alone (e.g. Collins and Evans 2007:31)

The SIH stresses the philosophical nature of the concept: under the right circumstances an individual can *fully* understand the world through the medium of language alone. As a bold conjecture, the SIH may be wrong. The initial empirical foundation for it is Sacks's account of 'Madeleine' (see above) and this is not a reliable source because Sacks had a different purpose in mind. The SIH is meant to unsettle thinking and encourage more experiments and observations – pressing thought and experiment to its limits is the job of a bold conjecture.

### *The fieldwork tributary*

The fieldwork tributary arose out of Collins's deep immersion, beginning the mid-1990s, in the society of gravitational wave physicists. Towards the end of the 1990s he noticed that his interactions with his respondents involved lots of physics talk of the technical kind which physicists had with each other. He called this ability to talk physics without doing it, 'interactional expertise'. Both streams of thinking are related to the separation principle. In the first stream this is exemplified by the situation of the congenitally disabled who cannot practice; in the second stream it is exemplified by the sociological fieldworker immersed in a technical domain who does not practice.

## *The imitation game tributary*

The imitation game (IG) tributary also arises out of AI. Two chapters of Collins, 1990 discuss the protocol of the Turing Test, arguing that it is a test of cultural embedding, not intelligence:

> A Turing-type test is a good test for computing ability precisely because it is a test of the extent to which a machine can be located in an interactional network without strain. Instead of asking about the innate ability of the machine one looks at its interactive competence; this is how we judge other things that interact with us. ... [they] fit neatly into our socio-cognitive networks (Collins 1990:190). [The other things being referred to are humans.]

Imitation Games are Turing Test with humans.[7] First experiments intended to explore the notion of interactional expertise were carried out on the colour blind and persons with perfect pitch (Collins and Evans 2014). We then tested Collins's – and various others' – ability to pass as gravitational wave physicists (Giles 2006). We then went on to test those who had been blind from a very early age. In each case, just as one would expect, the persons who has spent most time in embedded in the discourse of the other group were best at pretending to be that group: colour-blind person were better at pretending to be colour-perceivers than colour-perceivers were at pretending to be colour-blind; in contrast, 'pitch-perceivers' (those with perfect pitch) were better at pretending to be 'pitch-blind' (the rest of us) than the pitch-blind were at pretending to be pitch-perceivers; and, most strikingly of all, by a ratio of around 85/15, the blind were better at pretending to be sighted than the sighted were at pretending to be blind. After this, following Evans's use of the game in his sociology lectures to look at straights and gays and Christians and non-Christians, we were awarded a large grant to look at such topics cross-nationally (Collins and Evans 2014).[8] The imitation game continues to be used by various groups on a large and small scale to look at many different kinds of cultural difference.

---

[7] The Turing Test (Turing 1950) was, of course, based on an parlour game in which men pretended to be women and vice-versa

To repeat, the mainstream and the tributaries all originate from the sociological interpretation of Wittgenstein and that is why they all fit together. The philosophical stream disaggregates forms-of-life into practice and language components; the fieldwork tributary arises out of the focus of attention on the extent to which the analyst has acquired the native form-of-life; the imitation game tributary is to do with seeing the Turing Test as a measure of how well an individual (or computer) is embedded in a form-of-life.

***The sociology and policy stream***

Some wider implications of the idea of interactional expertise are shown in the sociology and policy column of Table 1. From the top, the idea illuminates the extent to which non-experts, such as political activists concerned with scientific and technological developments, can acquire enough understanding of the science and technology to make a sound contribution to debates (see also Collins 2014 and Collins and Evans 2015 and Collins, Evans and Weinel, 2016, forthcoming). The second entry refers to a thought experiment in Collins, 2004a, which tries to work out the technical 'rights' of a (special) interactional expert in case it should be discovered that gravitational waves were dangerous to health.[9] The following entries refer to the realisation that interactional expertise is the main resource of managers of technical projects (see Collins and Sanders, 2007) and that it is also the main resource of peer reviewers, and the members of technical committees where it can be observed that calculations and experiments are not done though the results of calculations and experiments are discussed. The next entry – 'sports coaches etc' we will return to since it marks a clear difference with the approach of Dreyfus which is relevant to the second part of this paper on embodiment. The next two entries (special IE has already been mentioned in footnote 8), concern interactional expertise and interdisciplinary collaboration – so-called 'trading zones' (Collins et al 2007). The following two entries indicate the potentially much more extensive reach of the concept of interactional expertise; it is claimed that without it we cannot understand how technical specialists coordinate their actions since they cannot all practice each other's actions and that therefore we cannot understand complex division of labour without the concept. Finally, the concept may be needed to explain the workings of society as a whole where, say, men and women are, perforce, unable to undertake certain of each other's practices yet must understand these mutually unpractised practices well enough to work together (or not!). The final entry refers to our attempt to use the concept as the basis for a suggestion for a new kind of committee (The Owls) which, in disputed areas, would reach and grade the best currently available technical understanding and feed it to policy-makers.

---

[8] The grant is an Advanced Research Grant awarded by the European Research Council (#269463, IMGAME, 2011-2016, €2,26M)

[9] The concept of 'special interactional expert' (Collins, 2011), arises consequent on realisation that contributory experts are interactional experts too so the 'special' has to be added to indicate that the expert has no contributory expertise.

The above is a brief sketch of the notion of the sources of interactional expertise and its applications. The force of the idea should become still clearer in the following part of the paper which deals with its relationship to the philosophical analysis of embodiment.

## Part 2: The body and language

### *Where we agree*

The first thing to say is that in relationship to the traditional 'artificial intelligentsia' and their hype I am in complete agreement with Bert Dreyfus and in pretty good agreement with his critique of what he calls 'Good Old Fashioned AI' (GOFAI). GOFAI is based on the idea that, to use Minsky's term, we are just 'meat machines'; when we do a calculation we are doing with our brains what a pocket calculator does with its program and silicon chips; when we execute a physical movement such as hitting a ball with a bat, we are somehow using our senses to create a description of the position and movement of things in space and time and our brains are doing physics calculations that can generate instructions to our muscles that will move our body parts such that the bat will hit the ball. Both Dreyfus's phenomenology and an extrapolation from the Wittgensteinian/Winchian (W/W) starting point lead to the same position in respect of this notion: it is ridiculous. In the case of W/W it is ridiculous because of the rules regress: rules do not contain the rules for their own application so any attempt to build a system that runs on rules of this kind is going to need further rules to know how to apply them and the problem is where those further rules come from. Dreyfus made all this clear at least as early as 1967 in his 'Why computers must have bodies in order to be intelligent' and with what he called the 'framing' problem and I got there about twenty years later from a somewhat different starting point. Dreyfus, of course, is 'the man', and deservedly so, while I am an interloper in respect of this debate, but the differences between us has become more and more marked and, I think, is now quite interesting and productive.

### *Where we disagree*

The bottom line of our disagreement is that Dreyfus thinks that the crucial thing is the body while I think that the crucial thing is embedding in society. Once more, the position is quoted most clearly in an abstract of a paper:

> [N]o machine will pass a well-designed Turing Test unless we find some means of embedding it in lived social life. *We have no idea how to do this* but my argument, and all our evidence, suggests that it will not be a necessary condition that the machine have more than a minimal body. Exactly how minimal is still being worked out. (Collins, 2008, p. 309, stress added for current purposes)

Later in this paper I am going to worry about whether my claim that we have no idea how to embed a computer in society is as secure as it once was.

The paper quoted above is the last in a short but flourishing argument between me, putting forward that the idea of interactional expertise against the idea of embodiment and the embodiment position put forward by Dreyfus and his supporters.[10] Unfortunately, in the way of so much academic life, this debate flared up and died without either party shifting their position. While I am inclined to think that is because the other parties were not prepared to admit that I was right even though they must know it 'in their heart of hearts' there are two other possibilities: we are talking across each other and my arguments are too obscure to understand. Here I'll have another go at trying to eliminate those possibilities.

---

[10] Notably Evan Selinger. The argument which can be dated from the 1996a review, continued in Collins's contribution to Dreyfus's *Festschrift* under the terms 'social embodiment thesis; 'individual embodiment thesis' and 'minimal embodiment thesis' (Collins 2000) and further published discussions followed in, for example, Selinger (2003), Selinger and Mix, (2004), Collins (2004c), Selinger, Dreyfus and Collins (2007), Selinger, 2008 and Collins, 2008. Very roughly, the philosophers (Dreyfus and Selinger) argued that the crucial point was that Madeleine had a body with a sense of front, back and so forth and that is why Lenat was wrong; the sociologist (Collins), argued that the interesting thing was the fluency that Madeleine could achieve with only a 'minimal body'.

## *The asocial body in general*

I am not a well enough trained philosopher to engage with Dreyfus et al on their home ground. Thus the apogee of the embodiment idea is said by Bert Dreyfus to be Samuel Todes's 1960s PhD thesis published in 2001 as *Body and World*, but I find that book very difficult because of the way it sets itself within the philosophical canon not to mention its scholarly density, which is typical of a PhD.[11]

What is clear, however, even from my lowly position, is that Todes is concerned with the role of our bodies in our experience of the world. Thus our horizontal orientation is conditioned by the fact that we have a front which favours movement forward, eyes pointing forward and so on, and a much less accessible back, and the fact that 'our bodily self-movement is itself grounded in the vertical field of gravitation' (pxiv of the Introductory remarks by Piotr Hoffman). But Todes remarks: 'The reader is forewarned that the analyses presented in this study are not of our normal experience in its full complexity ... Thus, for example, for the purposes of this study of the human body as the *material* subject of the world, our experience is simplified by disregarding our experience of other human beings.' (p1, author's stress). Since I am very much an autodidact in this area, and in a many other areas of philosophy, I am relieved to find an online review of the book which also points out.

> But the book bypasses entirely the fundamental human experiences of sociality and language—instead one could read Todes thinking that humans are hermits working out the meaning and efficacy of their participation in the world. The kinds of insights later hermeneuts and constructionists offer—that the categories we use to make our

---

[11] Dreyfus mentions the importance of Todes as far back as his 1967 and reiterated his view of the central importance of Todes's book at the meeting where this paper was first presented; his favourable views are still clearer in his published introductory remarks to Todes's book.

experience know-able and habit-able are accessible human and cultural constructions—were not available to Todes (Tom Strong, 'Bodies and Thinking Motion', *Janus Head*, 7(2), 516-522. Copyright © 2004 by Trivium Publications, Amherst, NY. Accessed last on 15 December 2015 at http://www.janushead.org/7-2/Todes.pdf)

Todes explains that he assumes that his experience of the material world is the same as that of every other human: 'I make the commonsense assumption that I live in the same world with you, my reader; and that your body plays the same role, philosophically speaking, in your life as mine does in my life.' (p1) Todes goes on to say that he hopes to complete a second study of humans' relationships with other humans – this, however, was never to be completed and Todes died in 1994.

Thus, by the time we are only a couple of pages into Todes's book we can see that we are probably talking across each other since I am simply unable to imagine how one can deal with perception and conceptualisation without seeing *the* major component as having to do with living in a social world: as having to do with socialisation and differential socialisation. The method of Todes is phenomenological – that is, it is an analysis of human experience – and from this develops a theory of the way humans in general interact with the material world. I have nothing against this project nor do I have grounds to argue against it. I am happy to accept what is said about humans in general – but I am not acquainted with any 'humans in general' only humans who see the world in specific and varying ways consequent upon the societies in which they live.[12]

---

[12] Another source is the sociology of knowledge given a Wittgensteinian interpretation.

One of the marked differences between Dreyfus and I is, then, that I start with social beings and he does not.  Furthermore, it is only possible for him to erect Todes to the status of hero (and I think the same goes for erecting Merleau-Ponty and Heidegger to hero status), by ignoring the social dimension of our engagement with the world.  And this all makes sense chronologically because, as the reviewer quoted above remarks: 'The kinds of insights later hermeneuts and constructionists offer—that the categories we use to make our experience know-able and habit-able are cultural constructions—were not available' to any of them.

[Let me now put something in parenthesis.]  Only with the growth of the sociology of scientific knowledge (SSK) in the early/mid1970s did the full extent of the social influence on perception become clear.  When even science and mathematics, the last redoubt of the asocial, turned out to be creatures of their social setting, then the reach of a universal theory of human perception narrowed still further.  Looking back with this in mind I think I finally understand a long-past incident.  In November 1991 Dreyfus was kind enough to contribute to an 'author meets critics' session at the annual meeting of the Society for Social Studies of Science held, that year, in Cornell University.  Dreyfus criticised my newly published (1990) book, *Artificial Experts: Social Knowledge and Intelligent Machines* which I thought of, somewhat fearfully, as no more than a sociologised version of his 1972 book.  My fear was that Dreyfus would say there was nothing new in my book and I did think that his 1972 was so well done that I would find such a charge hard to answer.[13]  In fact he took the opposite approach, treating my book as sufficiently different to be well worth attacking, which relieved me mightily. What took me by surprise, however, was his vehement criticism of my

---

[13] His remarks and my response were published as: Hubert Dreyfus, 'Response to Collins, *Artificial Experts'*, *Social Studies of Science,* Vol. 22, No. 4 (November 1992), 717-26; Collins, H. M., (1992) `Hubert Dreyfus, Forms of Life, and a Simple Test For Machine Intelligence', *Social Studies of Science*, 22, 726-39

SSK-style social constructivism, which seemed to me at the time not particularly important in respect of the discussion of computers. The attack was aimed at the way I tried to explain why some domains could be computerised and some could not – a matter of whether or not we chose to digitise our social life (restrict ourselves to machine-like or mimeomorphic actions) in respect of a domain.[14] But at last I see that an asocial conception of the world is central to a Todes/Dreyfus style of phenomenological analysis, or even a Heidegger, Merleau-Ponty style, so making everything social, as I was doing, would be anathema. [End of parenthesis.]

Todes, to repeat, puts forward a general theory of humans' interaction with the world. What implications that can be drawn from a theory of humans in general? The implications 'on the table' – those we find ourselves arguing about – are the possibilities and limitations of artificial intelligence and the capacities of specific individuals with unusual bodies and the source of the abilities of specific groups of skilled individuals. Let me try to explain what I see as the problem with the aid of an analogy. Linguists, such as Chomsky and his academic supporters and rivals, try to develop theories of language. That is, they try to explain the basic features of language and they study the ways humans acquire language. If one accepts what they say – let us suppose one is convinced by Chomsky's arguments for the existence of a universal transformational grammar – one then knows something about the limits and possibilities of human language in general. But this understanding provides only limited help if the topic one is interested in the language speaking abilities of feral children or the development of French, or German, or Swahili. If unusual bodies/brains or natural languages are the problems then generative grammar does not help very much. In the same way, general theories such as that of Todes, can't help much when it comes to the skills of particular groups, such as tennis-players or surgeons, nor when it comes to the particular way individuals such as the blind and the deaf cope with the world. In fact the contrary is the case: in so far as one might think that the general theory had something to say about the blind it seems to be wrong; surely we would expect individual blind persons, for whom the difference between front and back is much less marked than it is for the sighted, to live in a different world, and yet they do not.

---

[14] The term machine-like actions is taken from the 1990 book and is contrasted with the rather weak 'regular actions' but in our 1998 book Kusch and I change the terms to mimeomophic and polimorphic actions.

Well, I say they do not and yet we can find a suggestion that they do in H. G. Wells's short story, written in 1904, 'The Country of the Blind'. This, I suggest, is a brilliant exercise, *avant la lettre*, in the phenomenological analysis of the body. Wells indicates that a *society* made up of the blind *would* live in a different world. Thus, in a country of the blind there would be far less distinction between inside and outside and between day and night because neither walls nor darkness would present an obstacle to ordinary perception, now working through the ears rather than the eyes. I would imagine that the difference between front and back would also be less marked. In Well's country of the blind everything is different because of this differential embodiment. But the Country of the Blind is a whole different society illustrating how the body, language and the world interact at the collective level. What is strking is that in the societies of the sighted which we inhabit, the blind as individuals seem pretty similar to the rest of us in the way their world is constituted so a Todes-like theory actually leads us in the wrong direction. And that is because Todes, and presumably everyone who bases their work on his ideas, cannot see the difference between how things work at the collective and individual level and fail to cope with the fact that the body is, at best, far less important when it belongs to an individual.

## *Talking across each other*

In casual conversations with Bert we often have a little exchange that goes something like this:

> Bert: You say you can come to understand a practical ability by immersion in the spoken discourse. Do you think that just by talking about it you can learn to become a surgeon?
>
> Harry: Of course not, I would no more say that than I would say that you can teach someone to ride a bike by talk alone.
>
> Bert: Then we have no disagreement.

I think that these little snatches of conversation involve our talking across each other and the locus of mutual misunderstanding is the word 'understand'.[15] I think Bert is taking as an implicit definition of 'understanding a practice' the ability to execute that practice.[16] I am taking as an explicit definition of 'understanding', being able to pass a maximally demanding Turing Test or, rather, Imitation Game, in which the person who understands cannot be distinguished from a contributory expert – one who does know how to practice – in a verbal test. As I have tried to explain in Part 1 of the paper, this is not a trivial criterion, it is not just

---

[15] For snatches of a general theory of how different communities talk across each other, see Collins and Reber 2013. Reber is the leading psychologist of 'implicit learning'; we exchanged around 600 emails trying to work out why a psychologist and a sociologist, on the face of it working on the same topic, have never felt the need to cite each other's work.

[16] This was actually the definition of understanding that I used in my very first (1974) published paper in which I showed that TEA-laser scientists could learn to build a laser only if they acquired the tacit knowledge and this involved social contact with successful TEA-laser builders. The key to my discovery was rejecting the models of information exchange implicit in the literature on information diffusion -- which were all about what people read and how they came to read what they read – and replacing them with a stronger criterion – people had not to be able to read but to do – actually build a working TEA-laser – before I counted them as being possessors of TEA-Laser building knowledge. Bert is working with this criterion whereas for the purpose of what is going on here I have shifted to another criterion.

'talking the talk' without being able to 'walk the walk' because in a maximally demanding Imitation Game the interrogator will ask questions that bear on practical understanding such as 'How does it feel when you are first learning to balance on your bike?' and 'What does it feel like at the moment you know you are going to fall off?'

We are, of course, in the world of thought experiment here because it is unlikely that anyone who has not ridden has ever been sufficiently immersed in the discourse of bike-riding to be able to answer those questions – indeed such a deep discourse may not exist – but, under the *bold conjecture* – the 'strong interactional hypothesis' – the discourse would exist and the person immersed in it would be able to answer the questions. Does that mean that the person who has been so immersed in the discourse knows how it feels to first learn to balance and to fall off? Does it mean that sufficient immersion in the discourse means that the person has experienced the same 'feels' (the philosophical term is 'qualia' I believe) as the person who has actually learned to ride and actually fallen off? I would say that 'no' is the answer. But nevertheless the person with sufficient interactional expertise can make all the same judgments about the practical expertise in question as the person who has actually experienced the 'feels' and you would not be able to tell who is who by asking them questions. And that is why interactional expertise can do the work for technical managers who have to make those judgments and can do the work for all of us where we have to interact with others in respect of experiences of theirs that we have not shared – and that is nearly all the time. And that is why human societies are so different to animal societies and that is why a Todes-like analysis that does not give language an absolutely crucial position is not going to get us where we need to get to if we want to understand human understanding.

The disagreement is, then, stark. Bert says:

> You may have mastered the way surgeons talk to each other but you don't understand surgery unless you can tell thousands of different cuts from each other and judge which is appropriate. In the domain of surgery no matter how well we can pass the word along we are just dumb. So is the sportscaster who can't tell a strike from a ball until the umpire has announced it.
>
> I think the upshot of this is that Harry has discovered a new way of showing how the Turing Test fails to test intelligence and also fails to test linguistic expertise. He has shown that just being able to pass the word along is an inauthentic use of language. (Selinger et al, 2007 p 737)

In fact Bert even goes on to say, fairly desperately as I see it, that since I do seem to understand GW physics without doing it:

> Gravitational wave physics may be an interesting exception. *Science in general is desituated* but even science normally requires skilled discriminations. (Selinger et al, 2007 p 737, stress added)

Note that as late as 2007 when this was published, Bert was still sticking to the idea that science is desituated, still refusing to recognise the much richer understanding of science we have had since the 1970s.

### *Models of language and the limits of interactional expertise*

Wondering if the root of the problem is different notions of language, in Table 2 I try to set out what I think is going on in as clear a way as possible. As I see it, Bert and his colleagues live in a world where row 2 is beyond conception or too dangerous to be countenanced.

|   | THE ACTIVITY | ASSOCIATED MODEL OF LANGUAGE |
|---|---|---|
| 1 | Practical ability (contributory expert) | No language involved: 'walking the walk' |
| 2 | Special interactional expertise which involves an understanding of practical ability to the point of being able to make sound practical judgments even if does not carry the bodily sensations associated with it. In principle an interactional expert is indistinguishable from a contributory expert in a Turing Test | Language is a practice which is tacit-knowledge laden and technical judgment depends on tacit knowledge that can be captured by language. This is 'walking the talk'. |
| 3 | Formal or technical knowledge | Language, when used carefully, is a set of propositions like theorems which can express some of the rules contained in formal representations of physical processes which are found in scientists' abstractions and may by usable in computerised 'micro-worlds'. This is using talk as an abstraction |
| 4 | 'Passing the word along' | Language used in this way is a tricky and misleading attempt to pretend to have real practical understanding: it's just 'talking the talk not walking the walk' |

*Table 2: Are different conceptions of language at the heart of the problem?*

Let me be clear about what I do not want to claim for interactional expertise, that is, for *special* interactional expertise. Since we now believe that the relationship between contributory expertise and interactional expertise is transitive so that all experts are at least *latent* interactional experts (except, perhaps, for the exceptional cases in the first member of the list below) the question devolves into the capacities or interactional experts who are not contributory experts. Note that not all *latent* interactional experts become full-blown interactional experts: a *latent* interactional expert may not be able to realise their interactional expertise if they do not have interactional or reflective ability (ie they are inarticulate).[17] We'll use a list of questions for absolute clarity:

1) *Is it possible, in principle, to pass a Turing Test/Imitation Game in respect of every possible practical action? M*y inclination is to say 'no'. There are sensations that are so surprising in their intensity that I think no amount of speaking about them could put a person in position to make the right judgments in respect of them without having experienced them. For example, I cannot imagine that talk could ever reproduce the condition for the experience of (a) the rare intense sexual desire for an individual associated with either love or infatuation and (b) the witnessing of the birth of one's own child and the early years of parenthood. The astonishing intensity of these experiences is, in each case, a thing that I cannot imagine being imaginatively reconstructed. If language is not enough then anything that approaches interactional expertise would not be enough. (The case of soldiers' experience of battle is discussed in Collins 2012.)

2) *These few exceptions apart, are there special interactional experts in respect of all practical accomplishments?* Here the answer is definitely 'no' because to become an interactional expert requires that (a) there is a well-developed body of language – a 'practice language' – corresponding to the practical activity and (b) the speakers of the practice language are willing to absorb into their community someone who does not take part in their practices. There are probably lots of cases where condition 'a' is not fulfilled and the extent to which it is fulfilled for ordinary activities will vary from society to society. The same kind of variation will apply to condition 'b' and here matters are complicated by secrecy and time constraints.

---

[17] This is worked out in Collins and Evans 2007.

3) *Is interactional expertise always maximal?* The trouble here is that one does not know what maximal interaction expertise means. Maximal interactional expertise belongs more to the bold conjecture than the real world. In the real world we are almost always dealing with less than maximal interactional expertise – certainly in the case of special interactional experts – and we mostly don't know how complete a case we are dealing with. But this is not fatal. It is certainly important to keep in mind that acquiring interactional expertise is a difficult and generally long-drawn-out matter and important that it does not become a 'lite' concept justifying any activist's opinion on anything about which they know a little.[18] But none of this makes interactional expertise any harder to deal with than any other expertise – we are always having to make judgments about how expert someone is, sometimes employing tests but mostly not – while the idea of interactional expertise, and the idea of the imitation game, and the bold conjecture, keep our thinking on the right lines. Interactional expertise has important practical implications but it is also important not to let practical difficulties cloud the difference it makes to the relative importance of the language and body as compared to the phenomenological approaches discussed here.

### *Main differences and the narrow gaze of the embodiment thesis*

Let us finish by summing up some of the differences just mentioned or implied.

1) *Collective and Individual and general and particular*: The influence on understanding of the body and of specific bodily activities, such as playing various sports, surgery and specialties within science and technology, has to be treated separately at the collective and individual levels. At the collective level a Dreyfus analysis is correct: the form and content of our practical abilities and understanding has much to do with the form of our bodies: the general limits are probably set by a Todes-like analysis of forward-backward, vertical, within gravity and so forth, but this applies only at the highest level – it applies to human beings in general. Nested within this highest level are questions about particular activities in which some humans engage and others do not. The particular forms are worked out by a more Heidegger/Merleau-Ponty inspired Dreyfus-like analysis of the way we interact with the material world. But the Dreyfus-like analysis of individual's understanding of these collective practical activities is wrong because individuals who have not practised those collective practices, either because they cannot – the disabled – or have not – non-tennis-players or non-participants in GW physics – can still understand those practices through being immersed deeply enough in the corresponding practice languages. If we do not accept this we cannot understand how the world works: we cannot understand how sports commentators and coaches who are not themselves top-level ex-players commentate well and coach well; (Dreyfus says they must always do it badly but he is obviously wrong); we cannot understand how managers manage in technical domains nor can we understand peer review or the workings of technical committees; we cannot understand complex division of labour; and we cannot understand how the societies we live in could possibly work.

---

[18] For a discussion of 'lite' concepts and how to put a 'floor' under interactional expertise see Collins and Evans 2015.

2) *What is the body*? Since we see that the wheelchair bound and the blind are not struck dumb by their disabilities but seem to speak much as we do a question arises as to what the famous body consists of? We have two models on the table. The phenomenological model in which humans can extrapolate in some mysterious way even if they have bits of their body removed but, unfortunately, no one seems interested in how much can be removed before extrapolation becomes impossible. Why not? The interactional expertise model takes it that the minimal body is requires only enough components to enable it to engage in deep linguistic discourse.  I used to think these components would include ears, larynx etc, but I now see that with prostheses (think of Steven Hawking) the minimal body must be pretty close to a 'brain in a vat' so long as it is *well enough connected to society*.  What is clear, is that the brain must be capable of handling natural language: that now seems to me to be the only requirement so long as we set aside what it means to be connected into society.

3) *Where an understanding of interactional expertise leads*:  The intensely disappointing thing about the view of humans that does not take account of linguistic socialisation is the extreme narrowing of the range of questions asked.  Take language into account and as we have begun to see, a whole new set of inquiries are opened up.

   a. Why can't animals, which have bodies that are in most respects similar to us and live in the same gravitational field, communicate with each other in a rich way?  The foxhound knows nothing of the Chihuahua, the lion nothing of the domestic cat.

   b. Why are feral children unable to learn language?

   c. Why do the deaf find it so difficult to acquire a high level of linguistic skill unless they are taught language in some special way from birth?

   d. As already mentioned, why are the blind, as opposed to the deaf, so fluent?

   e. As already mentioned, why are the congenitally physically disabled so fluent?

   f. As already mentioned, how come we can understand things in respect of which we do not exercise our bodies?

   All these things seem to need answers to do with language and the brain: the gross body is a constant in a, b and, c where linguistic ability and understanding is markedly different to the norm while it varies in d and e, and its use varies in f, and fluency and understanding are unchanged or in the case of f, hardly changed if other circumstances are right. Here we are using a classic scientific research method – holding the body constant while watching understandings change and changing the body while watching understanding remain the same! In the normal way only one conclusion would follow: the body cannot the crucial causal factor when it comes to understanding – not at the individual level, anyway.

The really frustrating thing is that these questions, which one would think would be burningly central to anyone who was interested in the role of the body in understanding, are simply ignored by the embodiment theorists.

### *Computers, socialization and the internet*

Let us return to the big question with which both Dreyfus's critique of AI and the idea of interactional expertise began – the view that computers need bodies to be intelligent and the counter-view that they can manage without bodies but they cannot manage without socialisation. It is hard to go forward on the embodiment side of this because we do not really know what a body consists of, Selinger drifting toward my position in saying it can be a minimal thing with minimal properties.[19]

But my social-embedding position is not as secure as I once thought it was either. I said in 2008 that we have no idea how to socialise a machine. The trouble is that a lot has changed since the 1990s, when this idea was first worked out. I expressed my 1990s understanding in a critique of the Chinese Room thought experiment. I argued that even if the Chinese Room worked as advertised it would represent only a frozen moment of lived language. Language, being the property of the community, is always changing, so even if the Chinese Room could work as advertised for a day or two its way of using language would soon become archaic unless its look-up tables were continually refreshed by persons who were immersed in the changing flux of lived language. I argued that the intelligence would lie with the persons refreshing the tables, not the tables; the Chinese Room was like an extremely complicated telephone mediating normal conversation, not via converting sound waves into electrical signals and back again, but via a complicated trick with paper and print. We do not think telephones are intelligent so we should not think the Chinese Room is intelligent – the intelligence lies with two lots of ordinarily socialised native speakers: there are the speakers downstream of the conversation asking the questions and there are the speakers upstream of the conversation continually feeding-in updated look-up tables.[20]

---

[19] See his contribution to Selinger at al 2007.

Now, the problem for me as a critic of AI is that we now have automated ways of continually updating the look-up tables – something that I did not foresee a decade or so back! What has happened is that at today's frontiers of AI computers are continually connected with the internet! Furthermore, these computers that are continually connected to the internet are much more successful than any previous form of AI. Google is something like a working Chinese Room because it does not provide a fixed set of questions and answers but a continually updated set, the updates being based on what members of the embedding society are currently interested in – this is how Google's 'page-rank' works. It draws on the intelligence of the community of internet users. Other successful programs – that is programs that are more successful than those of a few decades back – also draw on the internet. As I

---

[20] A useful place to see this argument set out is the aforementioned Collins and Reber, 2013

understand it, these programs include IBMs 'Watson' and voice-interaction programs such as 'Siri'.

On the one hand, I could count this as a triumph for my theory: 'I said that AI isn't going to work unless they build socialised computers and the new generation of "socialised" computers work much better than anything we have seen before so I was right.' On the other hand, I don't really believe this is proper socialisation – socialisation is far more mysterious than that. So that is the next job: work out the difference between Google and true socialisation. I am betting, that for all the reasons set out above, it won't have anything to do with fronts, backs and gravity.